\documentclass[10pt,aps,prd,floatfix,notitlepage,showpacs,nofootinbib,superscriptaddress,twocolumn,showkeys]{revtex4-1}
\usepackage{amssymb,amsmath}
\usepackage[colorlinks = true, linkcolor = blue, urlcolor  = blue,  citecolor = blue, anchorcolor = blue]{hyperref}
\usepackage{xcolor}
\usepackage{graphicx}

\begin{document}
\title{Cosmological perturbations in gravitational energy-momentum  complex
}
\author{Habib Abedi}
\email{h.abedi@ut.ac.ir}
\affiliation{Department of Physics, University of Tehran, North Kargar Avenue, 14399-55961 Tehran, Iran.}

\author{Amir M. Abbassi} 
\email{ amabasi@khayam.ut.ac.ir}
\affiliation{Department of Physics, University of Tehran, North Kargar Avenue, 14399-55961 Tehran, Iran.}

\author{Salvatore Capozziello}
\email{capozziello@na.infn.it}
\affiliation{Dipartimento di Fisica, Universit\`a di Napoli ''Federico II'', Via Cinthia, I-80126, Napoli, Italy,}
\affiliation{Istituto Nazionale di Fisica Nucleare (INFN), Sez. di Napoli, Via Cinthia, Napoli, Italy,}
\affiliation{ Tomsk State Pedagogical University, ul. Kievskaya, 60, 634061 Tomsk, Russia, }
\affiliation{ Laboratory for Theoretical Cosmology,
Tomsk State University of Control Systems and Radioelectronics (TUSUR), 634050 Tomsk, Russia. }

\begin{abstract}
Starting from  the  energy-momentum of matter and gravitational field in the framework of General Relativity and Teleparallel Gravity,  we obtain the 
energy-momentum complex
in flat FRW spacetime. We show that the   complex vanishes at background level considering the various prescriptions, that is  the Einstein, M{\o}ller, Landau-Lifshitz and Bergmann ones. On the other hand, at level of  linear cosmological  perturbations, the energy-momentum complex is different from zero and coincides  in the various prescriptions. Finally, we evaluate the gravitational energy  for 
  different cosmological epochs governed by non-relativistic matter, radiation, inflationary scalar fields and cosmological constant.
\end{abstract}

\date{\today}



\maketitle
\section{Introduction}
Since the beginning of  general relativity (GR),  gravitational energy and its localization in curved spacetime has been one of the most controversial concepts in dealing with gravitational field. 
In general, the local conservation equation of matter energy-momentum is not valid in curved spacetime, i.e. ${T_{\nu\phantom{\mu},\mu}^{\phantom{\nu}\mu}\neq 0}$.  To construct the local conservation equation,  gravitational energy-momentum contribution is required. 
Einstein~\cite{Einstein} was known as the first one
who suggested energy-momentum pseudotensor. 
Misner~et~al.~\cite{Misner} proved that only the energy of a spherical system is localizable. Cooperstock et al.~\cite{Cooperstock1, Cooperstock2} argued that if energy of a spherical system is localizable, then the energy of any system can be localized. Then,  Bondi~\cite{Bondi} proved that non-localized energy is not allowed in GR.
Following the Einstein pseudotensor,  other prescriptions were developed;
 e.g. by M{\o}ller~\cite{Moller1}, Landau-Lifshitz~\cite{Landau1},
Papapetrou~\cite{Papapetrou1}, Bergmann~\cite{Bergmann1}, Tolman~\cite{Tolman1}, Weinberg~\cite{Weinberg1}, and so on.
All the calculations, except for the M{\o}ller prescription, must be carried out in Cartesian coordinates,  and despite the other forms of energy-momentum, gravitational energy-momentum prescriptions  are non-tensorial.

After the studies on the energy-momentum localization by Virbhadra et al.~\cite{Virbhadra}, this problem took again much attention.
These Authors investigated various prescriptions and showed that they coincide in some cases.
Chang et al.~\cite{Chang} argued that energy-momentum complex (EMC) can be associated to different Hamiltonian terms. 
The  EMC 
is  described by an antisymmetric super-potential, i.e.
\begin{equation}
\tau_\mu^{\phantom{\mu} \nu} = {\cal H}_{\mu \phantom{\mu \alpha} ,\alpha}^{\phantom{\mu}[\nu \alpha]}.
\end{equation}
Consequently, the divergence of EMC vanishes, i.e. $\tau_{\mu \phantom{\mu} , \nu}^{\phantom{\mu} \nu}=0$. Furthermore,  an arbitrary antisymmetric quantity can be added to the super-potential ${\cal H}_{\mu \phantom{\mu \alpha}}^{\phantom{\mu}[\nu \alpha]}$; it does not affect the conservation of the EMC $\tau_\mu^{\phantom{\mu} \nu}$.
Several attempts have been made to evaluate the total energy of the universe. It was shown  that the  EMC results in vanishing energy for open and closed Friedmann-Robertson-Walker (FRW) universe~\cite{Rosen2, Vargas1, Johri, Faraoni}. Most of this considerations can be reported in other formulations of gravitational field as teleparallelism.

The teleparallel description of gravity was introduced by Einstein in 1928~\cite{Einstein2}.
Teleparallel Gravity, which is a theory equivalent to GR with the same field equations, that is the Teleparallel Equivalent General Relativity (TEGR) can be  used to define the gravitational energy~\cite{teleparallel}.
The TEGR formulation uses the Weitzenb\"{o}ck connection which leads to dynamics without curvature $ R $, but with torsion $ T $.
Analogous to GR, various energy-momentum prescriptions in TEGR have been proposed.
We have to  notice that the Teleparallel version of EMC leads to the same  results in many cases~\cite{Gad}.  
The energy of the universe was also evaluated in TEGR~\cite{Ulhoa2}.
The energy localization problem has also been  considered in
 modified theories of gravity. In Ref.~\cite{Multamaki}, it  is investigated the possibility to extend  the Landau-Lifshitz complex to $f(R)$ gravity for the Schwarzschild-de Sitter universe.
Other modified prescriptions were developed in  $f(T)$ theory~\cite{Abedi1, Ulhoa, Ganiou} and in general higher-order theories of gravity~\cite{Capozziello1, Capozziello2, Rept, Oiko}.

In this paper, we will consider the problem of gravitational energy-momentum complex for GR and TEGR in the case of cosmology.

This study is organized as follows; 
In section~\ref{EMC} we briefly review the
energy-momentum distribution due to matter (all non-gravitational fields) and fields including gravity.
Considering Friedmann-Robertson-Walker (FRW) flat background  spacetime in section~\ref{E_FRW},
we explore the energy density complexes.
We use various EMC in GR and TEGR, i.e. Einstein, M{\o}ller, Landau-Lifshitz and Bergmann prescriptions.
The cosmological principle asserts that the universe is homogeneous and isotropic on large scales. 
 However, at smaller scales,  perturbations should be considered.
Writing the energy expressions in linear perturbations, 
we show an interesting coincidence between the prescriptions. In our knowledge, such expression never appeared in literature.
Finally, we  obtain the energy  in various epochs  including inflation, radiation, non-relativistic matter and cosmological constant-dominated epochs.
We summarize the results in section~\ref{Conclusion}.

Throughout this work, we use different indices: the Greek indices, i.e.  $\mu, \nu, \alpha, \cdots =0,1,2,3 $, are adopted  for the spacetime; the capital Latin  indices, i.e. $A,B,C, \cdots = 0,1,2,3$, are for the  tangent spacetime;  for  their related spatial indices, we use $i,j,k, \cdots = 1,2,3$ and $a,b,c, \cdots = 1,2,3$, respectively. We set the Planck mass $M_{\rm pl}^2=1$ and the signature ${(-,+,+,+)}$.
\section{The energy-momentum complex}
\label{EMC}
EMC in Einstein, Landau-Lifshitz, Bergmann-Thompson and M{\o}ller prescriptions, in the framework of GR,  are shown in Table~\ref{GREM}.
\begin{table*}[ht]
	\begin{center}
		\setlength{\tabcolsep}{1.5em}
		\begin{tabular}{l l l }
			\hline \hline
			prescription &Energy-Momentum pseudotensor & conservation law
			\\  \hline
			Einstein~\cite{Moller1} & $  \Theta^\nu_\mu = \frac{1}{2} H^{\nu \alpha}_{\mu,\alpha} $ & $\partial_\nu \Theta^\nu_\mu =0$	\\
			&
			$ H^{\nu \alpha}_\mu = -H^{\alpha \nu}_\mu = \frac{g_{\mu \sigma}}{\sqrt{-g}} \left[-g \left( g^{\nu \sigma} g^{\alpha \gamma} - g^{\alpha \sigma} g^{\nu \gamma}\right)\right]_{,\gamma} $
			&\\
			Landau-Lifshitz~\cite{Landau1} &
			$ L^{\mu \nu} = \frac{1}{2} \lambda^{\mu \nu \alpha \gamma}_{,\alpha \gamma}$
			& $\partial_\nu L^{\mu \nu} = 0$
			\\ &
			$\lambda^{\mu \nu \alpha \gamma}= -\lambda^{\mu  \alpha \nu \gamma} = -g \left(g^{\mu \nu} g^{\alpha \gamma} - g^{\mu \alpha} g^{\nu \gamma} \right)_{,\gamma}$ &
			\\
			Bergmann-Thompson~\cite{Bergmann1} & 
			$ B^{\mu \nu} = \frac{1}{2} \beta^{\mu \nu \alpha}_{,\alpha}$
			& $\partial_\nu B^{\mu \nu}=0$
			\\ &
			$\beta^{\mu \nu \alpha} = g^{\mu \sigma} v^{\nu \alpha}_\sigma$,
			\\ & $v^{\nu \alpha}_\mu=- v^{\alpha \nu}_\mu = \frac{g_{\mu \delta}}{\sqrt{-g}} \left[-g\left(g^{\nu \delta} g^{\alpha \gamma} - g^{\alpha \delta} g^{\nu \gamma}\right)\right]_{,\gamma}$ & \\
			M{\o}ller~\cite{Moller1} &
			$ M^\nu_\mu =  \chi^{\nu \alpha}_{\mu,\alpha}$ 
			& $\partial_\nu M^\nu_\mu=0$
			\\ &
			$\chi^{\nu \alpha}_\mu = - \chi^{\alpha \nu}_\mu= \sqrt{-g} \left(g_{\mu \sigma,\delta} - g_{\mu \delta,\sigma}\right) g^{\nu \delta} g^{\alpha \sigma}$\\
			\hline
			\hline
		\end{tabular}
		\caption{Energy-Momentum prescriptions in GR.}
		\label{GREM}
	\end{center}
\end{table*}
In this table, 
 EMC is defined, in the  Einstein prescription, as
\begin{equation}
\Theta^\nu_\mu=\sqrt{-g} \left(t^\nu_\mu+T^\nu_\mu\right),
\end{equation}
where $t^\nu_\mu$ and $T^\nu_\mu$ are the energy-momentum related to the gravitation and matter, respectively. We can consider 
 the conformal transformation 
\begin{align}
\hat{g}_{\mu \nu} =& \Omega^2(x^\alpha) \, g_{\mu \nu},
\end{align}
where the hat denotes  quantities in the new frame
and $\Omega(x^\alpha)$ is the conformal factor.
One can obtain the effect of  conformal transformation on the EMC, e.g.
 Einstein  prescription transforms as follows:
\begin{align}
2 \hat{\Theta}^{\phantom{\nu}\mu}_\nu =&  2 \Omega^2  \Theta^{\phantom{\nu}\mu}_\nu  +H^{\mu \alpha}_\nu \left(\Omega^2\right)_{,\alpha} 
\nonumber \\&
+ \frac{\left(\Omega^2\right)_{,\alpha \beta}}{\Omega^2} g_{\nu \sigma} (-g) \left(g^{\mu \sigma} g^{\beta \alpha} -g^{\beta \sigma}g^{\mu \alpha}\right) \nonumber
\\ & 
+\frac{\left(\Omega^2\right)_{,\alpha }}{\Omega^2} \left[g_{\nu \sigma} (-g) \left(g^{\mu \sigma} g^{\beta \alpha} -g^{\beta \sigma}g^{\mu \alpha}\right)\right]_{,\beta}.
\end{align}
Then,  for time dependent conformal transformation ${\Omega=\Omega(t)}$, we have:
\begin{eqnarray} \label{CT}
\hat{\Theta}^{\phantom{0}0}_0 &=&  \Omega^2  \Theta^{\phantom{0}0}_0  \label{EC}.
\end{eqnarray}
In TEGR approach, the metric can be written as $g_{\mu \nu}= \eta_{AB} h^A_{\phantom{A}\mu} h^B_{\phantom{B}\nu}$, where the nontrivial vierbein fields $h^A_{\phantom{A}\mu}$ are dynamical fields. 
One can transform  $ h^A_{\phantom{A}\mu} $ by a local Lorentz transformation without affecting the metric.
Therefore, metric $g_{\mu\nu}$ does not completely describe the vierbein fields.
 The  Weitzenb\"ock connection is defined by:
\begin{equation}
\tilde{\Gamma}^{\alpha}_{\phantom{\alpha} \mu\nu}:=h_{A}^{\phantom{A} \alpha}\partial_{\nu}
h^{A}_{\phantom{A} \mu}=- h^{A}_{\phantom{A} \mu}\partial_{\nu} h_{A}^{\phantom{A} \alpha
} .
\end{equation} 
This connection leads to vanishing curvature tensor and non-vanishing torsion tensor.
The tensor and scalar torsion, respectively, are given by:
\begin{align}
	T^{\alpha}_{\phantom{\alpha} \mu\nu} &= h_{A}^{\phantom{A} \alpha} \left( \nabla_{\mu} h^{A}_{\phantom{A}\nu} -\nabla_{\nu} h^{A}_{\phantom{A} \mu} \right),
	\\
	T &= T^{\alpha}_{\phantom{\alpha} \mu\nu} S_{\phantom{\mu\nu} \alpha}^{\mu\nu},
\end{align}
where $ \nabla_{\mu} $ is  the covariant derivative defined by 
the Levi-Civita connection and;
\begin{align}
	S^{\alpha\mu\nu}=&\frac{1}{4} \left(T^{\alpha\mu\nu}+T^{\mu\alpha\nu}-T^{\nu\alpha\mu} \right)
	\nonumber \\ &
	-\frac{1}{2}\left(g^{\alpha
		\nu}T^{\beta\mu}_{\phantom{\beta \mu} \beta}-g^{\mu \alpha}T^{\beta \nu}_{\phantom{\beta \nu} \beta} \right)
\end{align}
is a skew-symmetric tensor in its  last two indices.
Analogous to EMC in GR, teleparallel versions  can be  developed, see Table~\ref{TeleEM}.
	\begin{table*}[th]
		\small
		\begin{center}
			\setlength{\tabcolsep}{1.5em}
			\begin{tabular}{l l l }
				\hline
				\hline
				prescription &Energy-Momentum pseudotensor & conservation law 
				\\ \hline
				Einstein~\cite{Vargas1} & $h E_\mu^{\phantom{\mu}\nu} =2 \, \partial_\lambda\left(h ‌S_\nu^{\phantom{\nu}\mu \lambda}\right)$ 
				&  $\partial_\nu \left(h E_\mu^{\phantom{\mu}\nu}\right)=0$
				\\
				Landau-Lifshitz~\cite{Vargas1} &
				$h L^{\mu \nu} = 2 \, \partial_\lambda \left( h^2 g^{\mu \beta} S_\beta^{\phantom{\beta}\nu \lambda}\right)$
				& $\partial_\nu \left(h L_\mu^{\phantom{\mu}\nu}\right)=0$
				\\
				Bergmann-Thompson~\cite{Vargas1} & 
				$h B^{\mu \nu} = 2 \, \partial_\lambda\left(g^{\mu \beta} h S_\beta^{\phantom{\beta}\nu \lambda}\right)$
				& $\partial_\nu \left(h B_\mu^{\phantom{\mu}\nu}\right)=0$ \\
				M{\o}ller~\cite{Mikhail} &$h M^\nu_\mu = \partial_\beta {\cal M}^{\nu \beta}_\mu$ 
				&
				$\partial_\nu \left(h M_\mu^{\phantom{\mu}\nu}\right)=0$
				\\ &
				${\cal M}^{\nu \beta}_\mu = -{\cal M}^{\beta \nu}_\mu = \frac{\sqrt{-g}}{16 \pi} \Upsilon^{\tau \nu \beta}_{\chi \rho \sigma} \left[\Phi^\rho g^{\sigma \chi} g_{\mu \tau} - g_{\tau \mu} \gamma^{\sigma \rho \chi} \right]$,			
				\\
				&
				\quad
				$ \Upsilon^{\tau \nu \beta}_{\chi \rho \sigma} = \delta^\tau_\chi g^{\nu \beta}_{\rho \sigma} + \delta^\tau_\rho g^{\nu \beta}_{\sigma \chi} - \delta^\tau_\sigma g^{\nu \beta}_{\chi \rho}$,
				\\
				&
				\quad $g^{\nu \beta}_{\rho \sigma} = \delta^\nu_\rho \delta^\beta_\sigma - \delta^\nu_\sigma \delta^\beta_\rho$, \quad $\gamma_{\mu \nu \beta} = h_{i\mu} h^i_{\nu;\beta}$, \quad $\Phi_\mu = \gamma^\rho_{\phantom{\rho}\mu \rho}$ \\
				\hline
				\hline
			\end{tabular}
			\caption{Energy-Momentum prescriptions in TEGR}
			\label{TeleEM}
		\end{center}
	\end{table*}
The vierbein fields under  conformal transformation change  as follows:
\begin{align}
\hat{h}^{A}_{\phantom{A} \mu} =& \Omega(x^\alpha) \, h^A_{\phantom{A}\mu}.
\end{align}
One can obtain the effect of  conformal transformation on EMC in TEGR.
 Under conformal transformation we also have
\begin{align}
\hat{T}^{\rho}_{\phantom{\rho}\mu \nu} =& T^{\rho}_{\phantom{\rho}\mu \nu} + \Omega^{-1} \left(\delta^\rho_\nu \, \partial_\mu \Omega - \delta^\rho_\mu \, \partial_\nu \Omega\right),
\\
\hat{S}^{\phantom{\rho}\mu \nu}_\rho =& \Omega^{-2} S^{\phantom{\rho}\mu \nu}_\rho + \Omega^{-3} \left(\delta^\mu_\rho \, \partial^\nu \Omega - \delta^\nu_\rho \, \partial^\mu \Omega\right).
\end{align}
Consequently,  we can relate the Einstein complex in teleparallel formulation in two frames by
\begin{align}
\hat{h}  \hat{E}_\nu^{\phantom{\nu}\mu} =& \Omega^2  h E_\nu^{\phantom{\nu}\mu}+ h S_\nu^{\phantom{\nu}\mu \sigma} \, \partial_\sigma \Omega^2
\nonumber \\&
 + \partial_\sigma \left[h\Omega \left(\delta^\mu_\nu \, \partial^\sigma \Omega - \delta_\nu^\sigma \, \partial^\mu \Omega \right)\right].
\end{align}
Then similar to Eq.~\eqref{EC},  we have 
$\hat{h}  \hat{E}_0^{\phantom{0}0} = \Omega^2  h E_0^{\phantom{0}0}$
 for time dependent conformal factor.
The same results are valid  for all considered prescriptions in GR and TEGR.
Notice that the conservation of EMC is due to antisymmetric property of super-potential, and this property is invariant under conformal transformation.
Then, in general for all these prescriptions in GR and TEGR, we can conclude
\begin{equation}
\partial_\mu \left(\hat{h} \hat{\tau}_\nu^{\phantom{\nu}\mu}\right) = 0.
\end{equation}
In obtaining the local conservation of EMC,  it is standard to use conservation of matter energy-momentum tensor, i.e. 
$ \nabla_\mu T_\nu^{\phantom{\nu}\mu} = 0 $. In general, this equation does not remain valid under conformal transformation~\cite{Abedi18}, i.e.
\begin{align}
\hat{\nabla}_\mu \hat{T}_\nu^{\phantom{\nu}\mu}&= \Omega^{-4} \left(\nabla_\mu T_\nu^{\phantom{\nu}\mu} - T_\mu^{\phantom{\mu}\mu} \frac{\partial_\nu \Omega}{\Omega} \right),
\nonumber\\
\nabla_\mu T_\nu^{\phantom{\nu}\mu}&= \Omega^4 \left(\hat{\nabla}_\mu \hat{T}_\nu^{\phantom{\nu}\mu}
+ \hat{T}_\mu^{\phantom{\mu}\mu} \frac{\partial_\nu \Omega}{\Omega} \right).
\end{align}
Therefore, for any non-trace-free energy-momentum tensor, 
$ \hat{\nabla}_\mu \hat{T}_\nu^{\phantom{\nu}\mu} \neq 0 $. Because of  this fact,  the conformal transformation affects both  gravitation and matter, and the conservation $ \partial_\mu \left(h \tau_\nu^{\phantom{\nu}\mu}\right) = 0 $  is valid in both frames.

\section{The Problem of Energy in the FRW spacetime}
\label{E_FRW}
Let us consider now a FRW background spacetime with the following line-element
\begin{equation}
{\rm d}s^2 = a^2(\tau) \, \left(- {\rm d}\tau^2 +  \delta_{ij} \, {\rm d}x^i \, {\rm d}x^j \right), \label{bmetric}
\end{equation}
where $a(\tau)$ is scale factor and $\tau$ is the conformal time related to cosmic time $t$ by
$ {\rm d}\tau = {\rm d}t / a $.
The gravitational energy is expected to vanish for a Minkowski spacetime.
Using Eq.~\eqref{EC},  the energy   also vanishes for FRW spacetimes.

The Lagrangians of GR and TEGR differ by a boundary term, therefore they give rise to the same field equations.  Considering the cosmological case, the background field equations are 
\begin{eqnarray}
3 {\cal H}^2 &=& \rho  a^2, 
\\
-6 {\cal H}^\prime &=&   (\rho +3p) a^2 .
\end{eqnarray}
where ${\cal H}:=a^\prime / a$ is the comoving Hubble parameter, the prime indicates derivatives with respect to the conformal time $\tau$; $\rho$ and $p$ are the energy and momentum densities (pressure)  of the matter fields, respectively.
It is  also useful to write the following equation
\begin{equation}
{\cal H}^\prime = -\frac{1}{2} (1+3 w) {\cal H}^2, \label{H1}
\end{equation}
where, $w := p/\rho$ is the equation of state. Integrating Eq.~\eqref{H1} 
for a constant $ w$, we get
\begin{equation}
{\cal H} =\frac{2}{(1+3w)\tau}.
\end{equation}
If we apply the various EMC prescriptions to the FRW background, the energy complexes vanishes in all cases.
We can define linear scalar perturbations on the flat FRW‌ spacetime, in  conformal Newtonian gauge, as 
\begin{equation}
{\rm d}s^2 = a^2(\tau) \, \left[-(1+2 \Psi) \, {\rm d}t^2 + (1-2\Phi) \, \delta_{ij} \, {\rm d}x^i \, {\rm d}x^j \right]
\label{PertLine},
\end{equation}
where $\Psi$ and $\Phi$ are the Bardeen potentials, that is the  first order  perturbations.
The vierbein fields related to the line element~\eqref{PertLine} are given by
\begin{eqnarray}
h^{0}_{\phantom{0}\mu} &=& a \delta^0_\mu (1+\Psi) + a \delta^i_\mu \, \partial_i \alpha,
\\
h^{a}_{\phantom{a}\mu} &=& a \delta^a_\mu (1-\Phi) + a \delta^i_\mu B^{\phantom{i}a}_i + a \delta^0_\mu \, \partial^a \alpha ,
\end{eqnarray}
where $\alpha$ and $B^{\phantom{i}a}_i$ are extra degrees of freedom and $\partial_i \partial_j B^{ij} = 0$~\cite{Chen,Wu, Li1}. Extra degrees of freedom do not appear in metric.  The inverse vierbein fields get the following form 
\begin{eqnarray}
h_{0}^{\phantom{0}\mu} &=& a^{-1} \delta_0^\mu (1-\Psi) - a^{-1} \delta_i^\mu \, \partial^i \alpha,
\\
h_{a}^{\phantom{a}\mu} &=& a^{-1} \delta_a^\mu (1+\Phi) - a^{-1} \delta_i^\mu B_{\phantom{j}a}^i -a^{-1} \delta_0^\mu \, \partial_a \alpha .
\end{eqnarray}
We have also $\sqrt{-g}=h = \det\left(h^{(A)}_{\phantom{(A)}\mu}\right) = a^4 (1+\Psi - 3\Phi)$. The energy complex in GR and TEGR for  line-element~\eqref{PertLine} are shown in table~\ref{Energy}.
\begin{table}[t!]
	\small
	\begin{center}
		\begin{tabular}{l c c }
			\hline
			\hline
			prescription & Energy  in GR \;\; & Energy in TEGR  \\  \hline 
			Einstein & $ \frac{2}{a^2} \triangle \Phi $  & $\frac{2}{a^2}\triangle \Phi$\\
			Landau-Lifshitz & $ \frac{2}{a^2 } \triangle \Phi $ &   $\frac{2}{a^2}\triangle \Phi$\\
			Bergmann-Thompson & $ \frac{2}{ a^2} \triangle \Phi $ &  $\frac{2}{a^2}\triangle \Phi$ \\
			M{\o}ller & $ \frac{2}{ a^2} \triangle \Psi $ &  $\frac{2}{a^2}\triangle \Phi$\\
			\hline
			\hline
		\end{tabular}
		\caption{Energy complexes up to first order perturbations around FRW spacetime in GR and TEGR.}
		\label{Energy}
	\end{center}
\end{table}
All prescriptions except for M{\o}ller (in GR) are related to the matter energy density in comoving gauge $\rho_{\rm m}^{\rm c}$, 
\begin{equation}
\frac{2}{a^2} \triangle \Phi =-\delta \rho_{\rm m}^{\rm c},
\end{equation}
where $\triangle := \delta^{ij} \partial_i \partial_j$.
Notice also that  the energy complexes are proportional to the Ricci scalar of 3-space, ${}^{(3)}R= 4\triangle \Phi /a^2$.
The scalar part of matter energy-momentum tensor is
\begin{equation}
\left(  T_{\mu}^{\phantom{\mu} \nu} \right) =
\left(
\begin{array}{cccc}
-(\rho  +\delta\rho) & -(\rho+ p) \, \partial_{i}\delta u \\(\rho + p) \, \partial_{i}\delta u & \delta_{j} ^{i} (p +\delta p)  +p \left( \pi_{,ij}-\frac{1}{3} \delta_{ij} \, \triangle \Pi \right)      \end{array}
\right)    ,
\end{equation}
where $\delta \rho $ and $\delta p$ are perturbations in energy density and pressure, respectively.  The terms $\delta u$ and $\Pi$ are the potential velocity and the scalar part of the anisotropic stress.
 At first order of perturbations, the Einstein equations are
\begin{align}
\label{E0}
6{\cal H} \left(\Phi^\prime + {\cal H} \Psi\right) - 2 \, \triangle \Phi &= -  a^2 \, \delta \rho ,
\\
2\partial_i \left(\Phi^\prime + {\cal H} \Psi \right) &=  a^2 \left(\rho + p \right) \, \partial_i\delta u, \label{E27}
\\ 
\Phi^{\prime \prime} + {\cal H} \left(\Psi^\prime + 2 \Phi^\prime \right) + \left( 2 {\cal H}^\prime + {\cal H}^2 \right) \Psi &
\nonumber \\  
+ \frac{1}{3} \triangle \left(\Psi - \Phi \right) &=  a^2 \, \delta p/2, \label{dp}
\\
\partial_i \partial_j  (\Phi - \Psi) &= a^2 p \partial_i \partial_j  \Pi .
\end{align}
The last equation results in (see also \cite{Koivisto} )
\begin{equation}
  \Phi - \Psi  = a^2 p  \Pi .
\end{equation}
M{\o}ller energy in GR and TEGR are related   by 
\begin{align}
M_0^0\Big|_{\text{GR}} -M_0^0\Big|_{\text{TEGR}}
= 2 p \, \triangle \Pi.
\end{align}
Therefore, GR version of M{\o}ller energy complex would get a different value in presence of anisotropic pressure.
Hereafter, we consider the anisotropic part of energy-momentum tensor vanishing,  i.e. $\Pi=0$, so $\Psi = \Phi$.
 All prescriptions are, therefore, proportional to $\triangle \Phi$. Using the field equations we can write
\begin{eqnarray}
\Phi^{\prime \prime} + 3(1+c_{\rm s}^2) {\cal H} \Phi^\prime + 3 (c_{\rm s}^2 -w) {\cal H} \Phi = c_{\rm s}^2 \, \triangle \Phi \label{E33} ,
\end{eqnarray}
where
$c_{\rm s}^2 = \bar{p}^\prime / \bar{\rho}^\prime$
is the speed of sound which,
for a perfect fluid, is
$c_{\rm s}^2 = w$.
 Considering adiabatic perturbations and a  constant equation of state, i.e. $ w= \text{constant} $, the evolution equation of the Bardeen potential, from Eq.~\eqref{E33}, gets the following form in Fourier space (we use  the same symbols $\Phi$ in Fourier space),
\begin{equation} \label{35}
\Phi^{\prime \prime} + 3 {\cal H} (1+w) \Phi^\prime + w k^2 \Phi =0.
\end{equation}
In what follows we  will evaluate explicitly the gravitational energy of the universe assuming   various matter contents. We shall use the following component of the EMC
\begin{equation}
 \delta \tau^0_0 = -\frac{2}{a^2} \triangle \Phi \label{E35} .
\end{equation}
Then, using Eq.~\eqref{E0}, the gravitational energy  becomes
\begin{equation}
 \delta t^0_0 =  \delta \tau^0_0 - \delta T^0_0  = -\frac{6{\cal H}}{a^2} \left(\Phi^\prime+{\cal H}\Phi \right) =  -6H\left(\dot{\Phi}+ H \Phi \right),
\end{equation}
where $H = \dot{a}/a$ and dot refers to the derivative with respect to cosmic time $t$.
\subsection{Non-relativistic matter dominated epoch}
\begin{center}
 \begin{figure}[t!]
 		\includegraphics[width=3in]{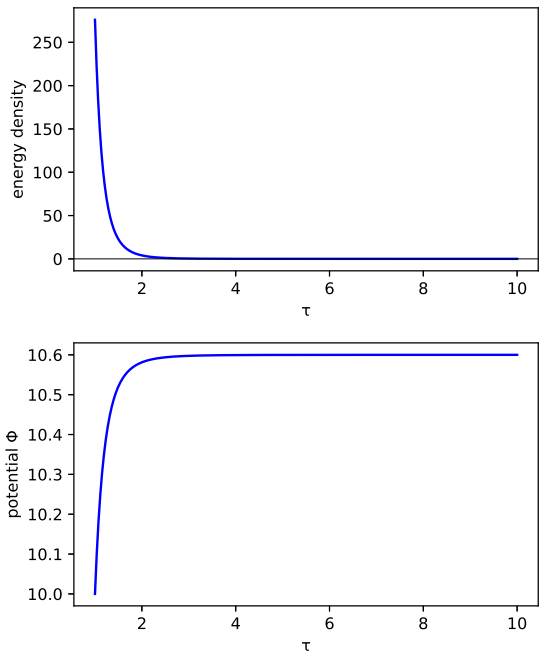}
 		\caption{The gravitational energy density and Bardeen potential  in matter dominated epoch.}
 		\label{fig:matter}
 \end{figure}
 	\end{center}

Consider the universe is dominated by pressureless, non-relativistic matter. The background matter energy density is 
$ \rho\propto a^{-3} $, then the scale factor and the comoving Hubble expansion rate become
\begin{align}
a(\tau) \propto \tau^2 &\propto t^{2/3}, 
&
{\cal H} &=\frac{2}{\tau}.
\end{align}
 Considering the background quantities, Eq.~\eqref{35}  reduces to
\begin{equation}
\Phi^{\prime \prime}+ \frac{6}{\tau} \Phi^\prime =0.
\end{equation}
This equation is scale  independent. 
 The solution is of the form
 \begin{equation}
 \Phi(\tau, {\bf x}) = c_1({\bf x}) + c_2({\bf x}) \, \tau^{-5},
 \end{equation}
 where $c_1$ and $c_2$ are integration constants that can be written by the initial value of the potential.
 As we can see,  the decaying mode disappears rapidly and the potential gets a time-independent value. 
 Finally, we have
 \begin{equation}
\delta t^{0}_0 =  -\frac{3 \tau_0^4}{a_0 \tau^6} \left(c_1 - \frac{3c_2}{2 \tau^5}\right) .
 \end{equation}
 Fig.~\ref{fig:matter} shows evolution of the gravitational energy and Bardeen potential  in matter dominated universe.
 
 \subsection{Radiation dominated epoch}
\begin{center}
 \begin{figure}[t!]
		\includegraphics[width=3.3in]{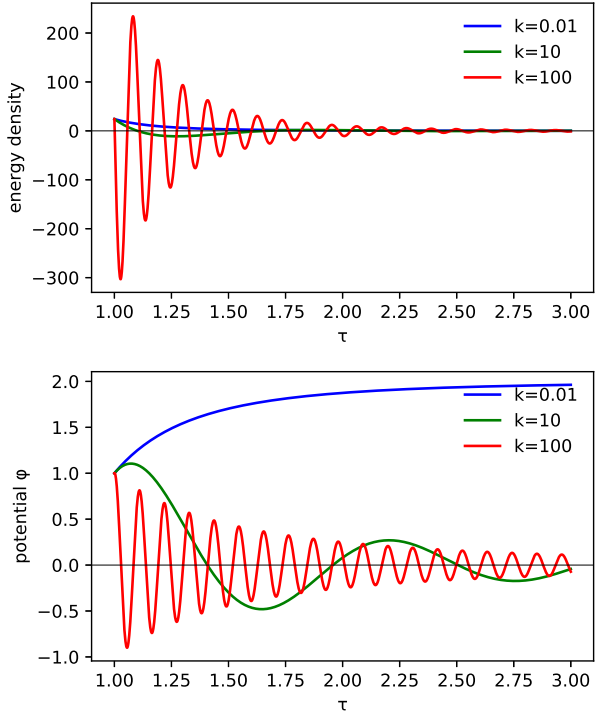}
		\caption{The gravitational energy density and Bardeen potential in radiation dominated epoch for various scales. }
		\label{fig:rad}
\end{figure}
\end{center}

 Now if we consider radiation dominated era with $\rho \propto a^{-4}$, we get
\begin{align}
a(\tau) &\propto \tau \propto t^{1/2},
&
{\cal H}&= \frac{1}{\tau}.
\end{align}
Then Eq.~\eqref{35} gets the following form
\begin{equation} \label{E43}
\Phi^{\prime \prime}+ \frac{4}{\tau} \Phi^\prime + \frac{1}{3} k^2 \Phi =0.
\end{equation}
Despite the matter dominated epoch,
Eq.~\eqref{E43} depends on scales ($k$-dependence). Defining $u := \tau \Phi$, the potential Eq.~\eqref{E43} gets the form of a spherical Bessel equation,
\begin{equation}
\frac{{\rm d}^2u}{{\rm d}y^2} + \frac{2}{y} \frac{{\rm d} u }{{\rm d} y} + \left[1-\frac{l(l+1)}{y^2}\right]=0,
\end{equation}
where $l=1$ and $y=k \tau / \sqrt{3}$.
Then the solution is 
\begin{equation}
u(\tau,{\bf x}) =c_1({\bf x})\, \,  j_1\left(\frac{k\tau}{\sqrt{3}}\right) + c_2({\bf x}) \, n_1 \left(\frac{k\tau}{\sqrt{3}}\right) ,
\end{equation}
where
\begin{eqnarray}
j_1(x) = \frac{\sin x}{x^2} - \frac{\cos x}{x} ,
\\
n_1(x) = -\frac{\cos x}{x^2}-\frac{\sin x}{x}.
\end{eqnarray}
The function $n_1(x)$ diverges for small $x$, therefore $c_2 =0$. The Bardeen potential gets the following form:
\begin{equation}
\Phi(\tau,{\bf x}) = \frac{9\sqrt{3}\, \Phi_0({\bf x})}{k^3 \tau^3}\left[\sin\left(\frac{k \tau}{\sqrt{3}}\right) - \frac{k \tau}{\sqrt{3}} \cos\left(\frac{k\tau}{\sqrt{3}}\right)\right].
\end{equation}
The potential, as mentioned before, depends on the scale.  For scales inside the horizon, i.e.
$k \tau \gg 1$, the potential is reduced to
\begin{equation}
\Phi(\tau,{\bf x}) = -9 \, \Phi_0({\bf x}) \frac{ \cos\left(\frac{k \tau}{\sqrt{3}}\right)}{\left(k \tau \right)^2} ,
\end{equation}
that is oscillatory with a decaying amplitude.
For superhorizon scales (${k \eta \ll 1}$), the potential becomes constant.
Finally, the  gravitational energy density becomes
\begin{align}
\delta t^{\phantom{0}0}_0 &= -\frac{6 \Phi_0}{\tau^4} \propto \frac{1}{a^4},
&
 k \tau& \ll 1,
 \\
\delta t^{\phantom{0}0}_0 &= -\frac{18\sqrt{3} \Phi_0}{ k \tau^4} \sin\left(\frac{k \tau}{\sqrt{3}}\right),
&
k \tau& \gg 1.
\end{align}
Fig.~\ref{fig:rad} illustrates the behavior of the gravitational energy density and the Bardeen potential in radiation dominated epoch for various scales.  At small scales, the gravitational energy and the Bardeen potential oscillate with decaying amplitude. At large scales, while the gravitational energy decays fast, the Bardeen potential tends to a constant value.

\subsection{Radiation and matter}
\begin{center}
	\begin{figure}[t!]
		\includegraphics[width=3.5in]{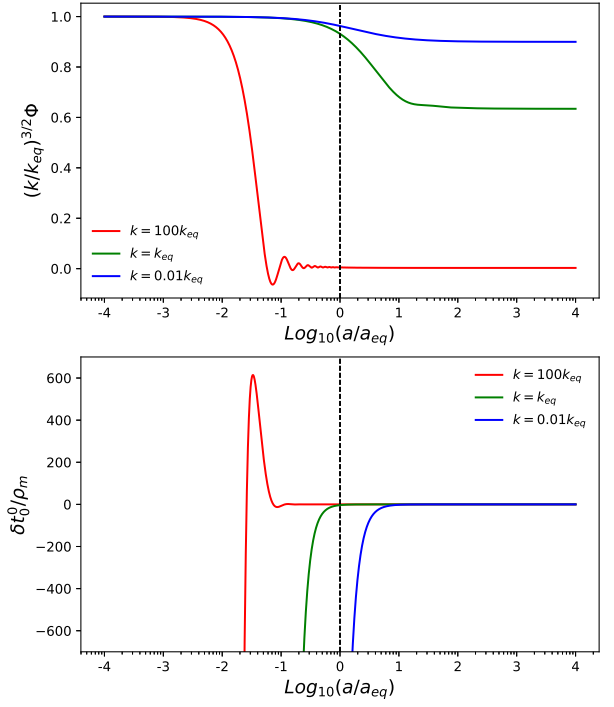}
		\caption{Evolution of the Bardeen potential and the gravitational energy density  with respect to  $\log_{10}\left( a/a_{\rm eq} \right)$,  for  universe containing radiation and non-relativistic matter.  Modes with $k<k_{\rm eq}$ and $k>k_{\rm eq}$ enter the Hubble radius  during matter- and radiation-dominated epochs, respectively. The vertical line at $a=a_{\rm eq}$ indicates the matter-radiation equality.}
		\label{fig:rm}
	\end{figure}
\end{center}

In the real universe there is no single component fluid. To be more realistic, it is convenient
to consider  radiation and non-relativistic  matter at the same time. 
Here we assume
 matter and radiation  non-interacting, therefore 
the conservation equation for each fluid in the background leads to:
\begin{align}
\rho_{\rm m} \propto& a^{-3}, 
&
\rho_{\rm r} \propto& a^{-4}.
\end{align}
The expansion is different in radiation and matter components.  However, there is an equivalence epoch where  the energy of matter and radiation  are comparable, therefore we  can define $t_{\rm eq}$ ‌as the  time when $\rho_{\rm m}=\rho_{\rm r}$ and use the scale factor as:
\begin{align}
 y = \frac{a}{a_{\rm eq}} = \frac{\rho_{\rm m}}{\rho_{\rm r}},
\end{align}
where radiation-matter equality occurs at $y=1$.
The Friedmann  equation results in the following scale factor
\begin{align}
y =& 2d \tau + d \tau^2,
\end{align}
where $d=\sqrt{\frac{\rho_{{\rm r},0}}{12}} \frac{a_0^2}{a_{\rm eq}} $.
The comoving Hubble expansion rate can be recast as follows
\begin{align}
{\cal H}(y) =& {\cal H}_{\rm eq} \frac{\sqrt{1+y}}{\sqrt{2} y},
\end{align}
where ${\cal H}_{\rm eq} = {\cal H}(y=1)$.
We assumed that there is no interaction between two components, however they affect each other gravitationally. Therefore, ignoring collision terms, covariant
energy-momentum tensor conservation is valid for each components. For linear perturbations, the conservation equation  results in the quantities:
\begin{align} \label{E56}
\delta_{\rm m}^\prime =& \triangle v_{\rm m} + 3 \Phi^\prime ,
\\ 
\label{E57}
v_{\rm m}^\prime = & - {\cal H} v_{\rm m}+ \Phi ,
\\
\label{E58}
\frac{3}{4} \delta_{\rm r}^\prime = & \triangle v_{\rm r} + 3 \Phi^\prime ,
\\
\label{E59}
v_{\rm r}^\prime =& \frac{1}{4} \delta_{\rm r} +\Phi ,
\end{align}
where $\delta_{\rm i}$ and $v_{\rm i}$ are the energy contrast and the velocity perturbation for the  component i.
Using Eqs.~\eqref{E56}-\eqref{E59}, the evolution equations of $\delta_{\rm m}(y)$ and $\delta_{\rm r}(y)$ can be written in term of scale factor $y$ in Fourier space,
\begin{align}
(1+y) \frac{{\rm d}^2 \delta_{\rm m}}{{\rm d} y^2} + \frac{2+3y}{2y} \frac{{\rm d} \delta_{\rm m}}{{\rm d} y}  = & 3(1+y)\frac{{\rm d}^2 \Phi}{{\rm d} y^2} 
\nonumber \\&
+ \frac{3(2+3y)}{2y} \frac{{\rm d}\Phi}{{\rm d} y} - \frac{k^2}{k_c^2} \Phi , \label{dy1}
\\
(1+y) \frac{{\rm d}^2 \delta_{\rm r}}{{\rm d} y^2}+\frac{1}{2} \frac{{\rm d} \delta_{\rm r}}{{\rm d} y} +\frac{1}{3} \frac{k^2}{k_c^2} \delta_{\rm r} = & 4(1+y) \frac{{\rm d}^2 \Phi}{{\rm d} y^2} +2 \frac{{\rm d} \Phi}{{\rm d} y} 
\nonumber \\ &
- \frac{4}{3} \frac{k^2}{k_c^2} \Phi , \label{dy2}
\end{align}
where $k_{\rm c} = {\cal H}_{\rm eq} /\sqrt{2}= k_{\rm eq} /\sqrt{2}$. 
The right hand side of Eq.~\eqref{E0} contains the total energy density perturbation, therefore we have:
\begin{align}
y \frac{{\rm d} \Phi}{{\rm d} y} + \Phi + \frac{1}{3} \frac{k^2}{k_c^2} \frac{y^2}{1+y} \Phi = & -\frac{1}{2} \frac{y}{1+y} \left( \delta_{\rm m} + y^{-1} \delta_{\rm r}\right). \label{dy3}
\end{align}
The initial conditions are adiabatic, on  super-Hubble scales  and in radiation dominated epoch; in other words $y\ll 1$ and $k\rightarrow 0$. At this time,  the potential $\Phi$ is time independent and 
$\delta_{\rm r} = -2\Phi$, $\delta_{\rm m} = -3\Phi/2 $,  $\delta_{\rm r}^\prime = 4\Phi^\prime$ and $\delta_{\rm m}^\prime =3 \Phi^\prime$.
Having the initial condition, it is straightforward to solve Eqs.~\eqref{dy1}, \eqref{dy2} and \eqref{dy3},  numerically. Fig.~\ref{fig:rm} describes the evolution of the potential and gravitational energy density for modes that enter the horizon at various epochs. 

\subsection{Inflation driven by scalar fields}
\begin{figure}[t!]
	\begin{center}
		\includegraphics[width=3.2in]{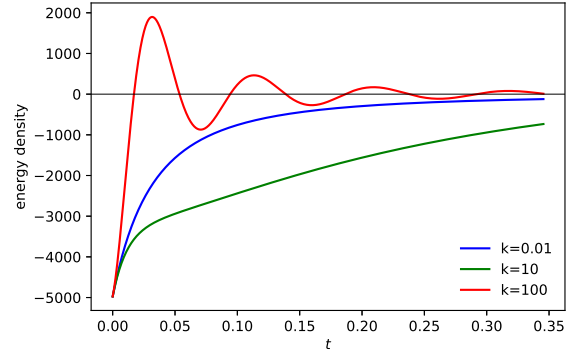}
	\caption{The gravitational energy in different scales for inflation driven by a scalar field $\phi$ with the potential ${V(\phi)=m_\phi^2 \phi^2/2}$, where $m_\phi \sim 10^{-6}m_{\rm pl}$. The initial conditions are considered as
	${\phi(0) = 3.2 m_{\rm pl}}$, ${  {\rm d}\phi(0)/ {\rm d}t=- 0.5 m_{\rm pl}}$, and as one expect for
	 the light scalar fields, we have used $\delta \phi(0) \sim  H$.
	}
		\label{fig:Inf}
	\end{center}
\end{figure}

Inflation not only provides convenient initial conditions  for big bang cosmology, but also provide the seeds for  structure formation. 
Quantum fluctuations, exiting the horizon, became classical and  are stretched due to exponential expansion. The  simplest dynamical candidate for describing inflation is achieved by a canonical scalar field.
Although the observations are in agreement with a single scalar field description of inflation,
the energy scales of inflation is beyond the standard model of particle physics. Theories formulated at these energy scales, in general,  contain several degrees of freedom that can be dealt under the standard of multiple scalar fields. 
In this section, we consider inflation driven by $N$ canonical scalar fields $\{\phi^i\}$, where $i = 1,2,3, \cdots , N$, minimally coupled to gravity.
The dynamics of background universe is described by:
\begin{align}
	{\phi^i}^{\prime \prime} + 2 {\cal H}{\phi^i}^{\prime} + a^2 V^{,i}&=0 , \label{2}
	\\
	 \sum_i \frac{1}{2} {\phi^i}^{\prime 2} +a^2 V(\phi^i) &=3{\cal H}^2 \label{3} ,
\end{align}
where $V(\phi^i)$ is a general potential and $V_{,i}:=\partial V / \partial \phi^i$.
The field equations in linear  perturbation regime become
\begin{align}
	{\delta \phi^i}^{\prime \prime} +2 {\cal H} {\delta \phi^i}^\prime -\triangle \delta \phi^i + \sum_j a^2 V^{,i}_{\phantom{,i}j}\,  \delta\phi^j  
	= -2a^2V^i \Phi 
	\nonumber \\
	+ 4 {\phi^i}^\prime \Phi^\prime, \label{4}
	\\
	\Phi^{\prime\prime}+3 {\cal H} \Phi^\prime+ a^2 V\Phi =\frac{1}{2} \sum_j \left({\phi_j}^\prime \, {\delta \phi^j}^\prime
	-  a^2V_j \, \delta \phi^j  \right) \label{5} .
\end{align}
The Poisson equation can also be written as follows:
\begin{align}
	2\triangle \Phi = \sum_j \left[{\phi_j}^\prime \, {\delta \phi^j}^\prime 
	- {\phi_j}^\prime {\phi^j}^\prime \Phi + 
	\left(3{\cal H}{\phi_j}^\prime+a^2 V_{,j}\right) \, \delta \phi^j \right]. \label{Poisson}
\end{align}
Also the $0i$-component of Einstein equation results in
\begin{align}
	\Phi^\prime = -{\cal H} \Phi + \frac{1}{2} \sum_j {\phi_j}^\prime \, \delta \phi^j. \label{0iEin}
\end{align}
For a given potential $V(\phi_i)$, one can solve Eqs.~\eqref{4}~and~\eqref{5} to obtain the evolution of $\Phi$ and consequently the gravitational energy density $\delta \rho_{\rm g} = -\delta t^0_0$, numerically.  Fig.~\ref{fig:Inf} Illustrates the gravitational energy density for a single scalar field model.
We use Eqs.~\eqref{Poisson}~and~\eqref{0iEin}  as constraint equations  for  initial values. For multi-field inflation we have
\begin{align}
	\delta t^0_0 =-3 \frac{\cal H}{a^2} \sum_j {\phi_j}^\prime \, \delta \phi^j.\label{10}
\end{align}
Considering slow-roll regime we can also solve the equations analytically. The
background Eqs.~\eqref{2} and \eqref{3} for multi-field inflation are reduced to
\begin{align}
	3 H \dot{\phi}^i + V^{,i} \approx 0 ,
	\\
	3H^2 \approx V.
\end{align}
We can also write the potential $\Phi$ as follows:
\begin{equation}
	\Phi = \frac{1}{2H} \sum_i \dot{\phi}_i \, \delta \phi^i \label{s1}.
\end{equation}
Then, the energy density becomes
\begin{equation}
	\delta t_0^0 = \frac{1}{2}\sum_i V_{,i} \, \delta \phi^i.
\end{equation}
The energy density in slow-roll regime is proportional to perturbation in potential~$V(\phi^i)$. The first order perturbations in matter energy-momentum tensor lead to:
\begin{equation}
\delta T^0_0 = -\sum_i \left(\dot{\phi}_i \, \dot{\delta \phi}^i - \dot{\phi}_i \dot{\phi}^i \Phi+ V_{,i} \, \delta \phi^i\right).
\end{equation}
The total energy therefore gets the following form:
\begin{equation}
\delta \tau^0_0 = -\sum_i \left(\dot{\phi}_i \, \dot{\delta \phi}^i - \dot{\phi}_i \dot{\phi}^i \Phi+ \frac{1}{2} V_{,i} \, \delta \phi^i\right).
\end{equation}
Eq.~\eqref{4} on super-horizon $k \ll Ha$ can be written as:
\begin{equation}
3H \, \dot{\delta \phi}_i + V_{,ij} \, \delta \phi^j \approx -2V_{,i} \Phi \label{s2}.
\end{equation}
For a given potential, one can solve  the Eqs.~\eqref{s1}~and~\eqref{s2}
to obtain $\Phi$ and $\delta \phi^i$.
 Considering  the sum-separable potential~${V(\phi^i)= \sum_j U_j(\phi^j)}$, the Bardeen potential and  scalar field perturbations are~\cite{Polarski}:
\begin{align}
	\Phi &= -C_1 \frac{\dot{H}}{H^2} -H\frac{{\rm d}}{{\rm d}t}\left(\frac{\sum_j d_j U_j}{\sum_j U_j}\right), \label{s3}
	\\
	\frac{\delta\phi_i}{\dot{\phi}_i} &= \frac{C_1}{H} - 2H\left(\frac{\sum_j d_j  U_j}{\sum_j U_j}-d_j\right), \label{s4}
\end{align}
where $C_1$ and $d_i$ are integration constants. Being the fields perturbations  in slow-roll regime, the constants $C_1$ and $d_j$can be calculated from Eq.~\eqref{s4}.
Finally, the energy density becomes
\begin{equation}
	\delta t^0_0 = 6C_1 \dot{H} + 6H^3\frac{{\rm d}}{{\rm d}t}\left(\frac{\sum_j d_j U_j}{\sum_j U_j}\right).
\end{equation}
As an example, let us  consider a simple two-field  model with  potential
$V(\phi,\chi)=m_\phi^2\phi^2 /2 + m_\chi^2 \chi^2 /2$.
This model is known as double inflation for
${ m_\phi \ll m_\chi} $~\cite{Silk}.  In slow-roll regime, the Bardeen potential and the scalar field perturbations, adopting the Newtonian gauge, get the following form
\begin{align}
\Phi = & -C_1 \frac{\dot{H}}{H^2} + \frac{2}{3} C_3 \frac{\left(m_\chi^2-m_\phi^2\right) m_\chi^2 m_\phi^2 \chi^2 \phi^2}{\left(m_\chi^2 \chi^2 + m_\phi^2 \phi^2\right)^2},
\\
\frac{\delta \phi}{\dot{\phi}} = & C_1 H^{-1} -2C_3‌H \frac{m_\chi^2 \chi^2}{m_\chi^2 \chi^2 + m_\phi^2 \phi^2},
\\
 \frac{\delta \chi}{\dot{\chi}} =& C_1 H^{-1} + 2 C_3 H \frac{m_\phi^2 \phi^2}{m_\chi^2 \chi^2 + m_\phi^2 \phi^2},
\end{align}
where
\begin{align}
\epsilon =-\frac{\dot{H}}{H^2} = & 2\frac{m_\phi^4 \phi^2 + m_\chi^4 \chi^2}{\left(m_\phi^2 \phi^2 + m_\chi^2 \chi^2\right)^2}.
\end{align}
The terms with $C_1({\bf k})$ and $C_3({\bf k})$ correspond to the growing adiabatic and isocurvature modes of solution.
Considering  scalar fields at the horizon crossing, one can write the coefficients $C_1$ and $C_3$. Finally, the gravitational energy reads:
\begin{align}
&\delta t_0^0 =
\nonumber \\ &
 -2\frac{ 
C_1\left(m_\phi^4 \phi^2 +m_\chi^4 \chi^2 \right)
+\frac{C_3}{3} \left(m_\chi^2-m_\phi^2\right) m_\chi^2 m_\phi^2 \phi^2 \chi^2}{m_\phi^2 \phi^2 + m_\chi^2 \chi^2}\,,
\end{align}
where the contribution of the single scalar fields is evident.
\subsection{Cosmological constant dominated epoch}
\begin{figure}[t!] 	\begin{center} 		\includegraphics[width=3.5in]{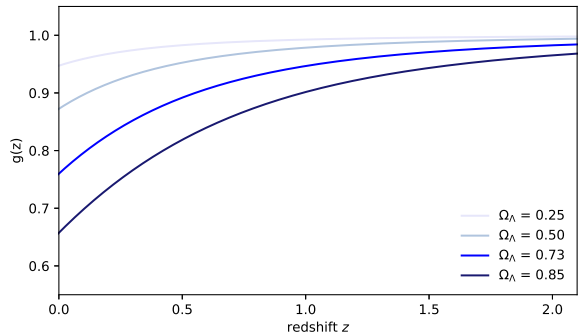} 	
		\caption{The suppression factor $g(z)=\Phi /\Phi_{\rm MD}$  for different values of density parameters: 
			${\Omega_\Lambda = 0.25 , 0.5 , 0.73 , 0.85}$. The blue line show the real $\Omega_\Lambda$.}		\label{fig:CC1} \end{center} \end{figure}

\begin{figure}[t!] 	\begin{center} 		\includegraphics[width=3.3in]{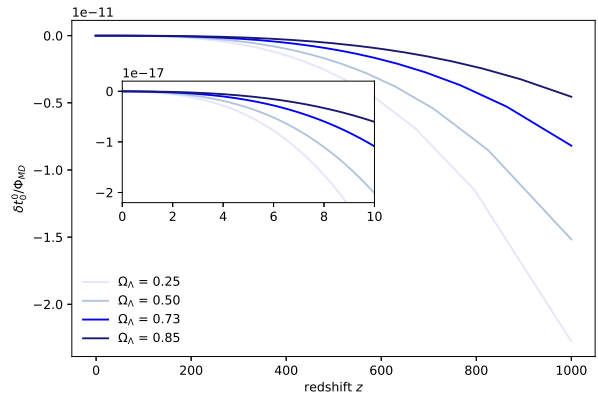} 	
 		\caption{ The gravitational energy density for different values of density parameters: 
 		${\Omega_\Lambda = 0.25 , 0.5 , 0.73 , 0.85}$.}		\label{fig:CC} \end{center} \end{figure}

In $\Lambda$-dominated epoch, the scale factor reads as
\begin{align}
a(\tau) =& -\frac{1}{H\tau},
&
{\cal H} =& -\frac{1}{\tau}, 
&
\tau <0,
\end{align}
where  the Hubble parameter $H$ is constant.
Here, the energy density of cosmological constant is obviously constant, i.e. $\delta \rho_\Lambda =0 =\delta p_\Lambda$. We consider cosmological constant in the presence of non-relativistic matter. Then the perturbations in matter may be important.
Eq.~\eqref{dp} in $\Lambda$-dominated epoch becomes:
\begin{align}
\Phi^{\prime\prime} - \frac{3}{\tau} \Phi^\prime + \frac{3}{\tau^2} \Phi =0.
\end{align}
The solution is
\begin{align}
\Phi = c_1 \tau + c_2 \tau^3.
\end{align}
Finally, the gravitational energy density becomes
\begin{align}
\delta t^0_0 = 12 c_2 H^2 \tau^3.
\end{align}
Notice that $\tau \propto {\rm e}^{-H t}$,
therefore, the energy decays exponentially.
To be more precise, we have to  consider the energy density of non-relativistic matter. From the Friedmann equation, we get the Hubble parameter,
\begin{align}
H = H_0 \sqrt{\Omega_{\rm m} (1+z)^3+\Omega_\Lambda},
\end{align}
where $z$ is the redshift. According to observations, we can assume 
$ \Omega_{\rm m} = 0.27 $ and $\Omega_\Lambda = 0.73$  the ratios of non-relativistic matter and cosmological constant energy density  to the critical energy density  at present epoch, respectively. We have ignored the energy density of radiation.
Eq.~\eqref{dp} is still valid in this case,
\begin{align} \label{E79}
\Phi^{\prime \prime} + 3\frac{a^\prime}{a} \Phi + \left( 2\frac{a^{\prime\prime}}{a} - \frac{{a^\prime}^2}{a^2} \right) \Phi = 0.
\end{align}
one can define the suppression factor $g(z)$ as follows
\begin{align}
\Phi = g(z) \, \Phi_{\rm MD},
\end{align}
where $\Phi_{\rm MD}$ is the potential   in matter dominated epoch.
We can solve Eq.~\eqref{E79} for $g(z)$ with the initial values $g(\infty)=0$ and ${\rm d}g(\infty) / {\rm d}z =0$. Having the function $g(z)$ we can plot the gravitational energy with respect to the redshift for different values of $\Omega_\Lambda$.
See Figs.~\ref{fig:CC1}~and~\ref{fig:CC} for $g(z)$ and $\delta t^0_0 / \Phi_{\rm MD}$.
\section{Conclusions}\label{Conclusion}
In present paper,  we have explored the issue of gravitational energy localization of  the universe in context of GR and TEGR. Considering fa lat FRW spacetime, we have applied different energy complexes assuming the prescriptions by  Einstein, M{\o}ller, Landau-Lifshitz and Bergmann. We found that the   energy complexes vanish at  background level. 
This result coincides with those in  previous works where the  Einstein and Landau-Lifshitz  prescriptions have been used  ~\cite{Rosen2, Vargas1, Johri, Faraoni}. Then we evaluated
the  gravitational energy of FRW spacetime considering cosmological  linear perturbations.  We found that the gravitational energy in all the  prescriptions considered of  this work  are identical and proportional to the matter-energy density in comoving gauge. Finally, 
we obtained the  gravitational energy for  
the universe filled with non-relativistic matter, radiation,   inflationary multiple scalar fields and cosmological constant.  In a forthcoming paper, considering results in \cite{Capozziello1,Capozziello2}, the same approach will be adopted in 
 the framework of modified gravity. The aim is to discriminate among  different modified theories of gravity considering cosmological EMC compared with observations. 
\begin{acknowledgments}
 SC is supported in part by the INFN sezione di Napoli, {\it iniziative specifiche} TEONGRAV and QGSKY.
The  article is also based upon work from COST action CA15117 (CANTATA),
supported by COST (European Cooperation in Science and Technology).
\end{acknowledgments}


\end{document}